\documentclass[journal,12pt,onecolumn,draftcls]{IEEEtran}
\ifCLASSINFOpdf
\else
\fi

\usepackage[T1]{fontenc}
\usepackage{color}
\usepackage{graphicx}
\usepackage{epstopdf}
\usepackage{caption}
\usepackage{subcaption}
\usepackage{cite}
\usepackage{amsmath}
\usepackage{amssymb}
\usepackage{algorithmic}
\usepackage{algorithm}
\usepackage{array}
\usepackage{fixltx2e}
\usepackage{float}
\usepackage{url}
\usepackage{multirow}
\usepackage{amsthm}

\begin{document}
%
\title{Joint Power Allocation and Beamformer for mmW-NOMA Downlink Systems by Deep Reinforcement Learning}
%
%
%

\author{Abbas~Akbarpour-Kasgari
        and~Mehrdad~Ardebilipour
\thanks{A. Akbarpour-Kasgari and M. Ardebilipour are with the Department
of Electrical and Computer Engineering, K. N. Toosi University of Technology, Tehran, Iran
e-mail: mehrdad@eetd.kntu.ac.ir.}
}

%
%

\markboth{Submitted to IEEE ~~~,~Vol.~, No.~, Date}%
{Akbarpour-Kasgari, and Ardebilipour \MakeLowercase{\textit{et al.}}: Joint power and beamformer for mmW-NOMA downlink}
%



\maketitle

\begin{abstract}
The high demand for data rate in the next generation of wireless communication could be ensured by Non-Orthogonal Multiple Access (NOMA) approach in the millimetre-wave (mmW) frequency band. Joint power allocation and beamforming of mmW-NOMA systems is mandatory which could be met by optimization approaches. To this end, we have exploited Deep Reinforcement Learning (DRL) approach due to policy generation leading to an optimized sum-rate of users. Actor-critic phenomena are utilized to measure the immediate reward and provide the new action to maximize the overall Q-value of the network. The immediate reward has been defined based on the summation of the rate of two users regarding the minimum guaranteed rate for each user and the sum of consumed power as the constraints. The simulation results represent the superiority of the proposed approach rather than the Time-Division Multiple Access (TDMA) and another NOMA optimized strategy in terms of sum-rate of users. 
\end{abstract}


%
\IEEEpeerreviewmaketitle


\section{Introduction}
%
%
%
%
\IEEEPARstart{M}{ultiple} access techniques due to their ability to provide suitable data accommodation, are the key enabler of the high data rate demands in future network generation. Among the well-known approaches, non-orthogonal multiple access (NOMA) is attracting researchers' interest to support high data rate demands and massive connectivity in 5G and beyond \cite{ChenZhang}. NOMA technique is introduced to outperform other well-known approaches such as time division multiple access (TDMA), orthogonal frequency division multiple access (OFDMA), and zero-forcing (ZF) in terms of spectral and energy efficiency \cite{ChenZhang}. NOMA allocates frequency and time resources between all the available users, simultaneously, thanks to the power domain superposition \cite{WangDaiWang}. 

The most crucial difference between mmW-NOMA and conventional NOMA is the interlace of power allocation and beamforming problems. In \cite{DingFan}, one of the pioneer works on mmW-NOMA schemes was adapted to introduce the random steering beamformer limited to the near users to each other. The energy efficiency was optimized subject to the power allocation and sub-channel assignment in \cite{FangZhang}, where no beamformer was represented. In \cite{HanifRatnarajah}, a fully digital beamformer was introduced to optimize the sum-rate. Moreover, in \cite{XiaoZhu}, the joint problem was considered but the optimization was only performed on the most powerful path in the channel. Further, in \cite{PangWu} joint power allocation and beamforming were optimized by considering Signal-to-Leakage-plus-Noise (SLNR) in order to decouple the power allocation and beamforming. On the contrary to all the mentioned prior works, we have used newly introduced machine learning (ML) tools to optimize the joint power allocation and beamformer design in mmW-NOMA. 

Recently, machine learning (ML) techniques are deployed in wireless communication networks as optimization tools. Among them, Reinforcement Learning (RL) is a closed-loop optimization procedure that is based on the action and reaction of the agent and environment, respectively. Further, encountering a continuous problem with a large number of states will be handled with deep RL (DRL) approaches \cite{MismarEvans, WangLi, GeLiang}. In \cite{MismarEvans} DRL is exploited to optimize joint beamforming, power control, and interference coordination in $5^{th}$ generation (5G) network. The authors used a deep Q-learning approach to estimate future rewards. Besides, Hybrid beamforming in mmW network was considered in multi-user multiple-input single-output (MU-MISO) system utilizing multi-agent DRL-based approach in \cite{WangLi}. Further, The optimal approach was designed based on employing a deep Q-learning approach and the cooperation of BSs in \cite{GeLiang}. 

In this paper, we have exploited DRL to optimize joint power allocation and beamformer in the Base Station (BS) to increase the sum-rate of two users. We have demonstrated joint power allocation and beamformer design in an optimization problem form. Further, we have demonstrated the agent and environment, the states, the actions, and the immediate reward to be maximized during the flow of the DRL algorithm. To handle the problem, we have borrowed actor-critic networks from DRL approaches with the aid of a deep neural network (DNN). Since the states of beamformer and power allocation are continuous, the DNN is employed to handle the complexity of the problem. Two different DNNs are exploited by actor and critic individually. The critic network accepts states of the environment and measures the action immediate reward. The actor network accepts states and the reward to generate the new action from space in order to maximize the future reward. 

The rest of the paper is organized as follows: Subsequent to the system model and problem formulation in section II, the proposed joint power allocation and beamformer design are represented in section III. Further, the numerical results and concluding remarks are demonstrated in section IV and section V, respectively. 



\section{System Model and Problem Formulation}
We consider a millimetre-wave non-orthogonal multiple access (mmW-NOMA) downlink system where an $N$-antenna equipped Base Station (BS) transmits towards $2$ single antenna users. In essence, the BS exploits $N$ antenna and NOMA strategy to simultaneously serves two users. Specifically, if we consider $s_k$ ($\mathbb{E}\left(|s_k|^2\right) = 1$) as the transmitted message for $k$-th user, the BS will transmit superposition $\sqrt{p_1}\mathbf{w}_1s_1 + \sqrt{p_2}\mathbf{w}_2s_2$ toward the users, where $p_k$, and $\mathbf{w}_k$ are the power devoted to the $k$-th user and the beamformer weight ($\| \mathbf{w}_k \|^2 = 1$), respectively. Accordingly, the received signal in the $k$-th user is denoted by 
\begin{equation}
y_k = \mathbf{h}_k^H \sqrt{p_k}\mathbf{w}_k s_k + \mathbf{h}^H_{\bar{k}} \sqrt{p_{\bar{k}}} \mathbf{w}_{\bar{k}} s_{\bar{k}} + n_k,
\end{equation}
where $\bar{k} = 2 (= 1)$ for $k = 1 (= 2)$, and $\mathbf{h}_k \in \mathbb{C}^N$ represents the complex channel vector between the BS and the $k$-th user. 

The channel vector between the BS and $k$-th user is an mmW channel. Since there are limited scatterers in the mmW band, the multipath which is caused by reflection is small and leads to a spatially sparse directional channel in the angle domain \cite{XiaoXia}. Assuming a uniform linear array (ULA), the mmW channel can be demonstrated as \cite{XiaoXia}
\begin{equation}
\mathbf{h}_k = \sum\limits_{l = 1}^{L_k}\lambda_{k,l}\mathbf{a}(N, \Omega_{k,l})
\end{equation}
where $\lambda_{k,l}$, and $\Omega_{k,l}$ are the complex coefficient and angle-of-arrival (AoD) of the $l$-th multipath components, respectively. Further, $L_k$ is the number of multipath components for $k$-th user, and $\mathbf{a}(.)$ is the steering vector function which is defined as 
\begin{equation}
\mathbf{a}(N, \Omega) = \left[ e^{j\pi 0 \cos(\Omega)}, e^{j\pi 1 \cos(\Omega)}, \dots, e^{j\pi (N-1) \cos(\Omega)}  \right]
\end{equation}
and is dependent to the geometry of the array. For the rest of the paper, we assume that the users are ordered based on their channel quality, i.e., $\| \mathbf{h}_1 \|_2 \leq \| \mathbf{h}_2 \|_2$. Exploiting Successive Interference Cancellation (SIC), the NOMA receiver can decode the power domain signal transmitted signal in the same frequency and time resource \cite{WunderJung}. Based on the suggested ordering the received Signal-to-Interference-plus-Noise (SINR) in each user can be defined as 
\begin{eqnarray}
\text{SINR}_1 = \frac{|\mathbf{h}^H_1 \mathbf{w}_1|^2 p_1}{|\mathbf{h}^H_1 \mathbf{w}_2|^2 p_2 + \sigma^2} \\
\text{SINR}_2 = \frac{|\mathbf{h}^H_2 \mathbf{w}_2|^2 p_2}{\sigma^2} 
\end{eqnarray}
where $\sigma_2$ is the noise variance. Hence, the achievable rate in each user is denoted as 
\begin{eqnarray}
R_k = \log_2(1 + \text{SINR}_k). 
\end{eqnarray}

In this paper, we consider the problem of joint power allocation and beamformer design for sum-rate optimization. As a consequence, the objective function is the sum-rate of two users. Moreover, there are some constraints which are denoted as below: 
\begin{equation}\label{eq.opt}
\begin{array}{rrclcl}
\max_{\mathbf{w}_k,p_k} & R_1 + R_2\\ 
s.t.  & R_k \geq r_k \\ 
 & p_1 + p_2 = P \\
 &  \| \mathbf{w}_k \|_2^2 = 1
\end{array}
\end{equation}
where $k = 1,2$ for the above equation and $P$ is the total available power in BS. Moreover, $r_k$ is the minimum achievable rate in $k$-th user which should be provided. As discussed earlier, the cost function is the sum-rate of the users. Further, the first constraint ensures the minimum rate for each user. Besides, the second and third constraints determine the feasible space of the independent variables of the problem, i.e., they guarantee the transmission power of the users and their corresponding beamforming weights. 



\section{Joint Power Allocation and Beamformer Design}
Here we will discuss the proposed approach to solve the joint power optimization and beamformer design problem in Eq. \eqref{eq.opt}. We have utilized the Deep Reinforcement Learning (DRL) approach to introduce the optimized power and beamformer weights. 

\subsection{Deep Reinforcement Learning}
DRL is involved of Deep Learning (DL) and Reinforcement Learning (RL), simultaneously. Due to existence of the DL part in this approach, it's very suitable for handling large state and action space Markov Decision Process (MDP) problems \cite{SuttonBarto}. The MDP can be defined by a four-tuple space as $\{\mathcal{S}, \mathcal{A}, \mathcal{P}, \mathcal{R}\}$, where $\mathcal{S}$ is the state space denoting the observation from environment. Further, $\mathcal{A}$ is the action space and representing the available actions from the agent to optimize the reward. Besides, $\mathcal{P}$ is the state transition probability, and $\mathcal{R}$ is the immediate reward of the action $a_t\in \mathcal{A}$ in state $s_t \in \mathcal{S}$. According to the immediate reward, the long-term reward $V(s) = \mathbb{E}\lbrace \sum\limits_{t=0}^{\infty}\gamma^t r_t(s_t,a_t) | s\rbrace$ for discount factor $\gamma \in [0,1]$ can be maximized to guarantee the best action in each state. 

Accordingly, we can model the problem at hand into the DRL framework by defining mandatory 4-tuple state variables. To represent the 4-tuple state variables, we demonstrate the immediate reward function based on the problem. To this end, we need to reconsider the optimization problem in Eq. \eqref{eq.opt}. The auxiliary variable $\alpha$ is defined to guarantee the feasibility of the constraint. Hence, the following immediate reward is defined 
\begin{equation}\label{eq.score}
	r_t = (1 - \alpha)(R_1 + R_2)
\end{equation}
where $\alpha = 1$ if one of the constraints is not met, while $\alpha = 0$ ensures the constraints feasibility. In other words, if each of the constraints in Eq. \eqref{eq.opt} is not met, then $\alpha = 1$. Consequently, immediate reward in the episode will be zero. 

Hence, the state $\mathcal{S}$ is defined as $\mathcal{S} = \lbrace \mathbf{h}_1^{\mathcal{R}}, \mathbf{h}_1^{\mathcal{I}}, \mathbf{h}_2^{\mathcal{R}}, \mathbf{h}_2^{\mathcal{I}}, \alpha \rbrace$, where superscript $\mathcal{R}$ and $\mathcal{I}$ denote real part and imaginary part, respectively. Furthermore, the action state $\mathcal{A}$ is defined as $\mathcal{A} = \lbrace \mathbf{w}_1^{\mathcal{R}}, \mathbf{w}_1^{\mathcal{I}}, \mathbf{w}_2^{\mathcal{R}}, \mathbf{w}_2^{\mathcal{I}}, p_1, p_2 \rbrace$. 
\subsection{Deep Deterministic Policy Gradient}
\begin{figure}
\includegraphics[width = \linewidth]{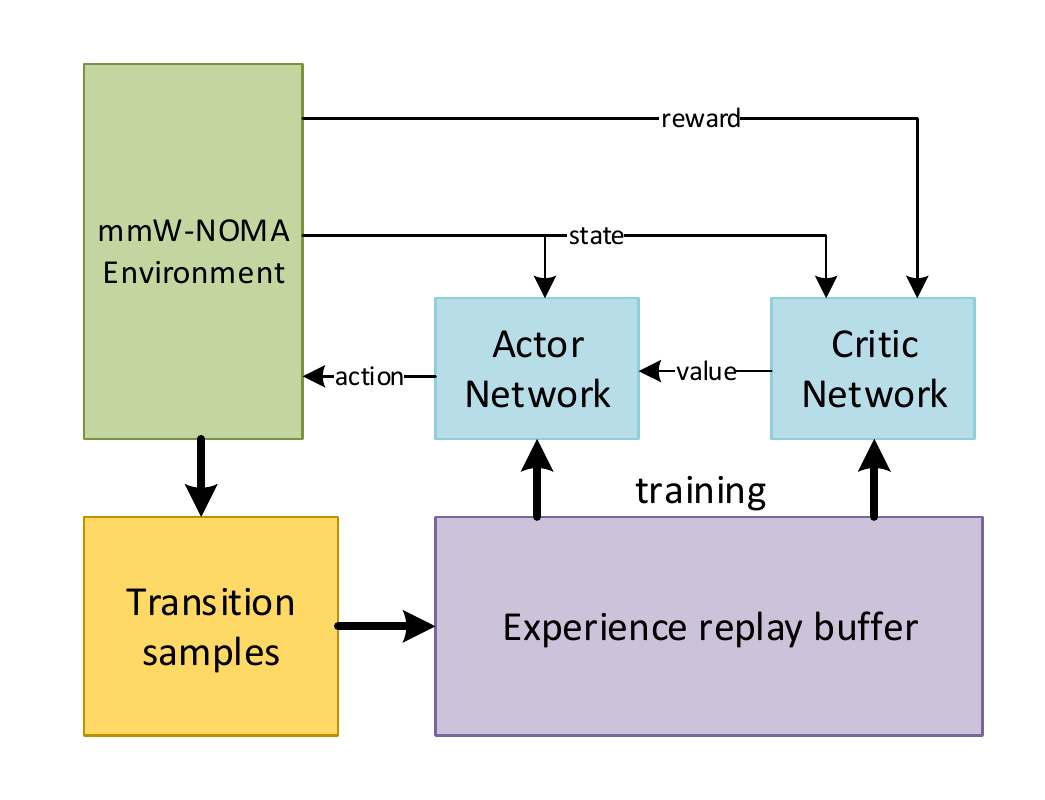}
\centering
\caption{Schematic diagram of actor-critic network}
\label{Fig.ActorCriticNet}
\end{figure}

Due to the continuous state space of the joint power allocation and beamformer, we have exploited the Deep Deterministic Policy Gradient (DDPG) approach to map the state $s_t \in S$ to the action space $a_t \in \mathcal{A}$ such that the value function is maximized. DDPG algorithm utilizes a DNN with the parameter $\theta$ to estimate the optimal policy function \cite{SuttonBarto}. The DNN training tries to update the policy $\pi_\theta $ in the gradient direction of value function defined as 
\begin{equation}\label{eq.grad}
	\triangledown_{\theta}J(v) = \mathbb{E}_{s}\left[ \triangledown_aQ(s,a|\beta) \triangledown_\theta \pi_\theta (s) |_{a=\pi_\theta(s)} \right] 
\end{equation}
where $J(v)$ is the value function and defined as
\begin{equation}
	J(v) = \sum_{s\in\mathcal{S}}p(s)\sum_{a \in \mathcal{A}} \pi_\theta (a|s) Q(s,a|\beta)
\end{equation}
The gradient formulation in Eq. \eqref{eq.grad} inspires the actor-critic framework where two separate DNN parameters corresponding to $(\theta, \beta)$ are optimized \cite{SuttonBarto}. The actor network will update $\theta$ and demonstrate the optimal policy, while the critic network will update $\beta$ to demonstrate the intermediate value function for the current action which has been applied by the agent. The whole network of the system is demonstrated as in Fig. \ref{Fig.ActorCriticNet}.

\subsection{Deep Neural Networks}
As mentioned earlier, the whole optimization network involved two independent DNNs. The first one which is the critic one accepts two different inputs including current observation and current action. These inputs are merged via a concatenation layer and then go across three dense layers followed by ReLU as the activation layer. Finally, a dense layer will estimate the Q-value of the network. 

Secondly, the actor network welcomes the current action and after applying four dense layers with ReLU as the activation layer, will determine the next action of the agent.

\section{Numerical Results}
In this section, the simulation results are shown. Further, the results are compared with the TDMA approach which is denoted as 'TDMA' in the figures. Moreover, the NLOS system in \cite{XiaoZhu} is considered as another benchmark for comparison which is called 'NLOS-NOMA' in the figures. The simulation results are averaged on $1000$ independent runs. In the DDPG part, the forgetting factor, actor learning rate, and critic learning rate are $0.99$, $0.0001$, and $0.0005$, respectively. Further, the number of episodes is $1000$ in training while the number of runs over each episode is $250$. The average score is defined based on the last $250$ score. Additionally, the number of users is two, while the number of transmitter and receiver antennas is $16$. The signal-to-noise ratio (SNR) is defined as the ratio of total power to the noise power ($P/\sigma^2$). 

\begin{figure}
	\includegraphics[width=0.99\linewidth]{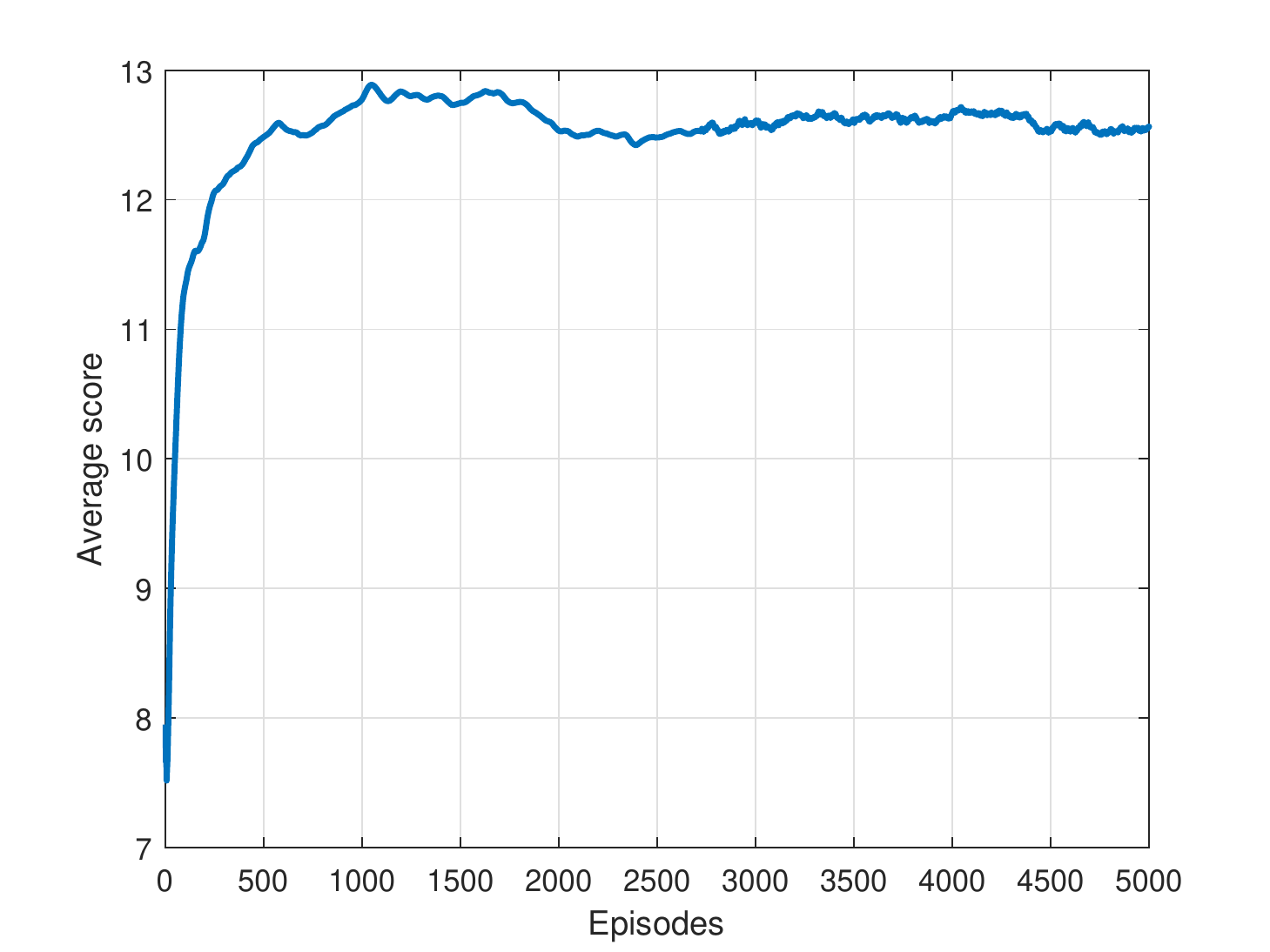}
    \caption{The average score value during training in 5000 episodes}\label{fig.score}
\end{figure}

The score of the training part is represented in Fig. \ref{fig.score}. As depicted, $5000$ episodes are used for training. Each episode is consist of $250$ trials and the score is moving average of the last $250$ consecutive results. Moreover, the SNR is $30$ dB, and $r_1 = r_2 = 1$ (bps/Hz). As represented, the achievable sum-rate is increased from $7$ to more than $12$ and stabled around it. This simulation represented that the algorithm is converged. 

\begin{figure}
	\includegraphics[width=0.99\linewidth]{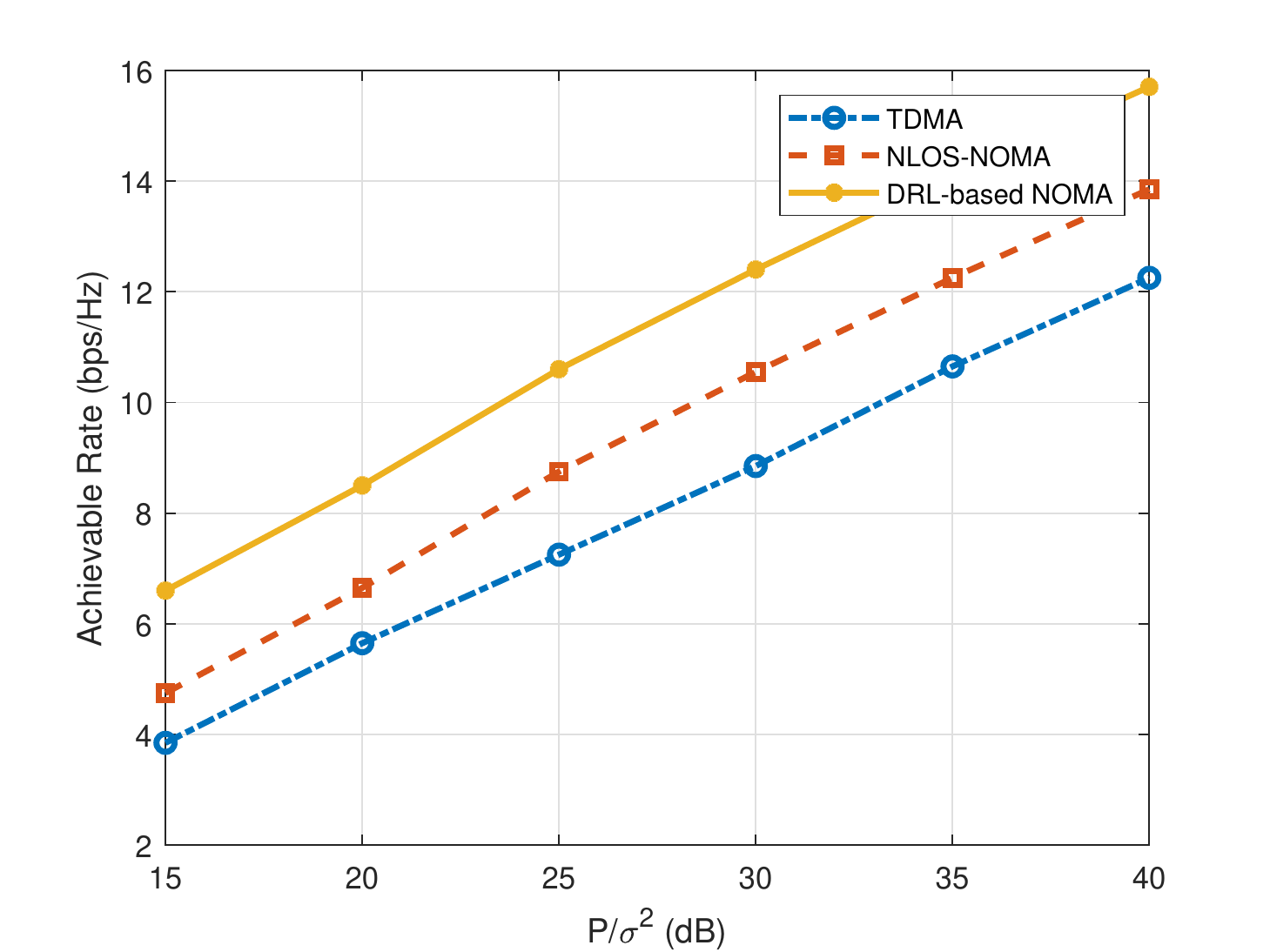}
	  \caption{The achievable sum-rate comparison regarding the SNR of signal}\label{fig.snr}
\end{figure}

In Fig. \ref{fig.snr}, the achievable sum-rate is compared for different algorithms in various SNRs. Here, the minimum guaranteed rate for each user is equal to $1$ (bps/Hz). As stated before, TDMA and NLOS-NOMA are utilized as the benchmarks of comparison. As can be seen, the sum-rate of the proposed DRL-based approach outperforms the other approaches. The superiority of the proposed approach to TDMA is obviously due to the simultaneous resource allocation in NOMA rather than TDMA. Further, the proposed DRL-based approach is superior to the NLOS-NOMA which could be two-fold. On one hand, the proposed approach tries to simultaneously optimize the power allocation and beamformer design, while in NLOS-NOMA the problem is broken into two individual parts. On the other hand, the proposed approach is not limited to the channel condition, while in NLOS-NOMA, the power of the LOS tap is very effective in the performance of the algorithm. 

\begin{figure}[htb]
\includegraphics[width=0.99\linewidth]{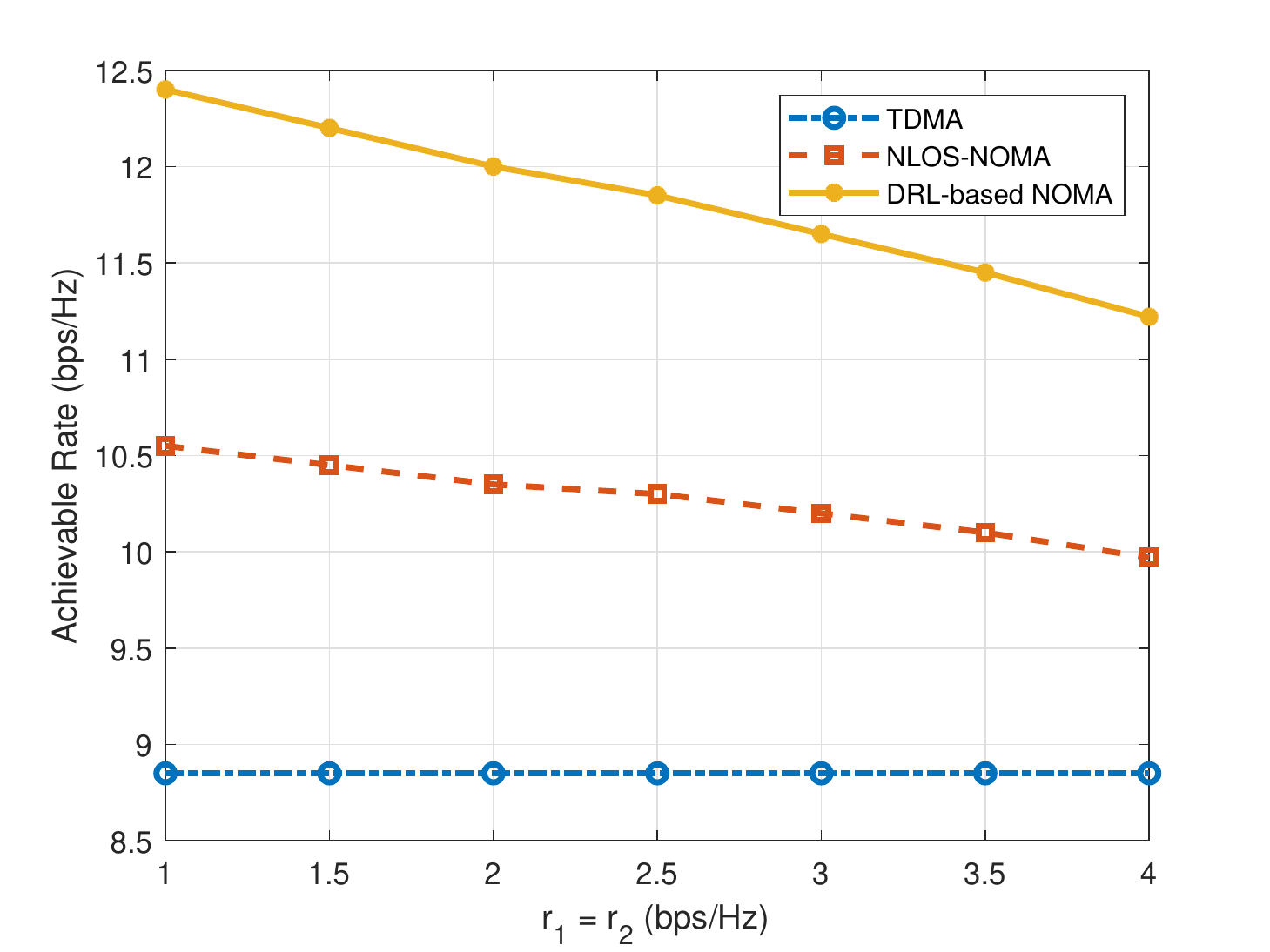}
  \caption{The achievable sum-rate comparison regarding the guaranteed minimum rate}\label{fig.rate}
\end{figure}

In Fig. \ref{fig.rate}, the impact of the minimum guaranteed rate for each user is considered on the achievable sum-rate with the predefined SNR = $30$ dB. Here, the proposed DRL-based approach is superior to the other benchmarks, too. With some more accurate consideration, it is obvious that the gap between the proposed DRL-based approach and the NLOS-NOMA approach is decreased with the increase of the minimum guaranteed rate. This is caused by the score definition. In Eq. \eqref{eq.score}, we have defined $\alpha = 0$ or $\alpha = 1$. In essence, we have defined a step that cut the score calculation very sharply if the conditions are not satisfied. This would increase the number of zero scores in higher minimum guaranteed rates. Consequently, the optimization problem is encountered with some discrete points which is not favorable. Hence, in our future works, we will consider a soft limiter in the score calculation and represent its performance. 

\section{Concluding Remarks}
In this paper, we have considered the joint power allocation and beamformer design in the mmW-NOMA downlink system. In order to optimize the problem, we have exploited a well-known data-science-borrowed DRL approach with the actor-critic framework. The most crucial advantage of the proposed approach rather than previous ones is the joint optimization while in other previous works the joint problem is broken into two individual ones due to non-convexity. We have defined action space, states, and immediate reward of the system based on the sum-rate of two users, as well. Ultimately, we have compared our proposed DRL-based approach with TDMA and NLOS-NOMA based approaches as the state-of-the-art algorithms. As represented in the figures, the proposed approach outperforms other algorithms perfectly, due to simultaneous optimization and limitless behavior encountering various channels. 

\section*{Acknowledgement}
The authors would like to thank from Iran National Science Foundation (INSF) which supported this work financially under grant number 99011213.
\ifCLASSOPTIONcaptionsoff
  \newpage
\fi

\end{document}